\documentclass[preprint,review,11pt,authoryear]{elsarticle}

\usepackage{amsmath}
\usepackage{amssymb}
\usepackage{amsthm}
\usepackage{array}
\usepackage{eurosym}
\usepackage{graphicx}
\usepackage{lipsum}
\usepackage{multirow}
\usepackage{subcaption}
\usepackage{xcolor}
\usepackage{color}
\usepackage{soul}
\usepackage{url}

\newcommand{\yh}{\hat{y}}
\newcommand{\xb}{\boldsymbol{x}}
\newcommand{\betab}{\boldsymbol{\beta}}
\newcommand{\thetab}{\boldsymbol{\theta}}
\newcommand{\tabitem}{~~\llap{\textbullet}~~}

\usepackage[left=2.4cm, right=2.4cm, top=2.4cm, bottom=2.4cm]{geometry}

\journal{European Journal of Operational Research}

\begin{document}
	
\begin{frontmatter}
		

\title{Instance-Dependent Cost-Sensitive Learning \\for Detecting Transfer Fraud}

\author[MathematicsAddress]{Sebastiaan H\"oppner}
\ead{sebastiaan.hoppner@kuleuven.be}
\author[EconomicsAddress,SouthamptonAddress]{Bart Baesens\corref{correspondingauthor}}
\ead{bart.baesens@kuleuven.be}
\author[VUBAddress]{Wouter Verbeke}
\ead{Wouter.Verbeke@vub.be}
\author[UAAddress,MathematicsAddress]{Tim Verdonck\corref{correspondingauthor}}
\cortext[correspondingauthor]{Corresponding author: Tim Verdonck}
\ead{Tim.Verdonck@uantwerpen.be}

\address[MathematicsAddress]{KU Leuven, Department of Mathematics, Celestijnenlaan 200B, Leuven 3001, Belgium}
\address[EconomicsAddress]{KU Leuven, Faculty of Economics and Business, Naamsestraat 69, Leuven 3000, Belgium}
\address[SouthamptonAddress]{University of Southampton, School of Management, Highfield Southampton, SO17 1BJ, United Kingdom}
\address[VUBAddress]{Vrije Universiteit Brussel, Faculty of Economic, Political and Social Sciences and Solvay Business School, Pleinlaan 2, B-1050 Brussels, Belgium}
\address[UAAddress]{University of Antwerp, Department of Mathematics, Middelheimlaan 1, Antwerp 2020, Belgium}

\begin{abstract}
Card transaction fraud is a growing problem affecting card holders worldwide. Financial institutions increasingly rely upon data-driven methods for developing fraud detection systems, which are able to automatically detect and block fraudulent transactions. From a machine learning perspective, the task of detecting fraudulent transactions is a binary classification problem. Classification models are commonly trained and evaluated in terms of statistical performance measures, such as likelihood and AUC, respectively. These measures, however, do not take into account the actual business objective, which is to minimize the financial losses due to fraud. Fraud detection is to be acknowledged as an instance-dependent cost-sensitive classification problem, where the costs due to misclassification vary between instances, and requiring adapted approaches for learning a classification model. In this article, an instance-dependent threshold is derived, based on the instance-dependent cost matrix for transfer fraud detection, that allows for making the optimal cost-based decision for each transaction. Two novel classifiers are presented, based on lasso-regularized logistic regression and gradient tree boosting, which directly minimize the proposed instance-dependent cost measure when learning a classification model. The proposed methods are implemented in the \texttt{R} packages \texttt{cslogit} and \texttt{csboost}, and compared against state-of-the-art methods on a publicly available data set from the machine learning competition website Kaggle and a proprietary card transaction data set. The results of the experiments highlight the potential of reducing fraud losses by adopting the proposed methods. 
\end{abstract}

\begin{keyword}
Decision analysis \sep	Fraud detection \sep Cost-based model evaluation \sep Cost-sensitive classification 
\end{keyword}

\end{frontmatter}


\section{Introduction}
\noindent

In September 2018 the European Central Bank issued the fifth oversight report on card fraud. The report analyses developments in fraud that are related to card payment schemes (CPSs) in the Single Euro Payments Area (SEPA). The report indicates that the total value of fraudulent transactions conducted using cards issued within SEPA and acquired worldwide amounted to \euro 1.8 billion in 2016. In relative terms, i.e. as a share of the total value of transactions, the total value of fraudulent transfers amounted to 0.041\% in 2016 \citep{fraudreport2018}. Therefore, developing powerful fraud detection systems is of crucial importance to financial institutions in order to reduce losses by timely blocking, containing and preventing fraudulent transactions. 

A stream of literature has reported upon the adoption of data-driven approaches for developing fraud detection systems \citep{phua2010, ngai2011}. Although these methods significantly improve the efficiency of fraud detection systems, opportunities exist to better align the development of data-driven fraud detection systems with the actual business objective. The objective that is adopted in learning from data by many data-driven approaches is \emph{statistical} in nature, e.g., the likelihood or cross-entropy is maximized, whereas a more appropriate objective in learning would be to minimize the losses due to fraud. 

For this purpose, cost-sensitive learning methods \citep{sahin2013} may be adopted, which can take into account class-dependent misclassification costs \citep{chan1998}. These methods are often adopted to address the class imbalance problem in fraud detection \citep{dalpozzolo2014}, since they allow to emphasize the importance of correctly identifying observations of the minority class, i.e. fraudulent transactions. 

Recently, a number of methods that take into account transaction- or instance-dependent misclassification costs have been proposed and evaluated for detecting credit card fraud \citep{bahnsen2014example, bahnsen2017fraud}. Alternative methods that aim at optimizing for the business objective while learning have been proposed that adopt a profit-driven strategy. These methods maximize the performance of the resulting fraud detection model as evaluated using a customized profit measure \citep{verbeke2012, hoppner2018}. 

In this paper, we further extend upon this recent stream of literature, by developing a theoretical underpinning for profit-driven, example-dependent learning, as well as by proposing and adopting a customized profit measure and objective function for application in developing a data-driven credit card fraud detection system. More specifically, we introduce two novel instance-dependent cost-sensitive methods: cslogit and csboost, which are adapted from logistic regression and gradient tree boosting. Both methods adopt an objective function which assesses both the profitability of a model, by means of customized profit measure, and the complexity of a model, by means of alasso penalty.  Both methods have been implemented as \textsf{R} packages, including plot, summary and predict functions and will be published so as to allow reproduction of the results presented in this paper (with the exception of the results on a proprietary dataset). Both methods can be applied to any classification problem involving an instance- or class-dependent cost-matrix.

This paper is organized as follows. In Sections 2 and 3, an instance-dependent cost-sensitive framework is introduced for making cost-optimal decisions with respect to card fraud detection. Sections 2 and 3 provide a theoretical underpinning for developing a customized objective function for data-driven learning in Section 4, as implemented within logistic regression and gradient tree boosting in Section 5. Section 6 further details the cslogit and csboost algorithms and describes the user interface of their implementation. Section 7 presents the results of an empirical evaluation of the proposed approaches. Finally, concluding remarks and potential directions for future research are provided in Section 8.

\section{Instance-dependent cost-sensitive framework for  transfer fraud detection}
	
The aim of detecting transfer fraud is to identify transactions with a high probability of being fraudulent. From the perspective of machine learning, the task of predicting the fraudulent nature of transactions can be presented as a binary classification problem where instances belong either to class $0$ or to class $1$. We follow the convention that the instances of interest such as fraudulent transactions, belong to class $1$, whereas the other instances such as legitimate transfers, correspond to class $0$.  We often speak of positive (class $1$) and negative (class $0$) instances.

In general, a classification exercise leads to a confusion matrix as shown in Table \ref{tab:confusion_matrix_general}. For example, the upper right cell contains the instances belonging to class $1$ (e.g. fraudulent transactions) which are incorrectly classified into class $0$ (e.g. predicted as being legitimate).
\bgroup
\def\arraystretch{1}
\setlength{\tabcolsep}{5pt}
\begin{table}[t]
	\begin{center}
		\begin{tabular}{ r c c }
			\hline
			& Actual legitimate & Actual fraudulent \\
			& (negative) $y = 0$ & (positive) $y=1$ \\\hline
			Predicted as legitimate & True negative & False negative  \\
			\vspace{2mm}
			(negative) $\yh = 0$ & $[C_i(0|0)=0]$ & $[C_i(0|1)=A_i]$ \\
			Predicted as fraudulent & False positive & True positive \\
			(positive) $\yh = 1$ & $[C_i(1|0)=c_f]$ & $[C_i(1|1)=c_f]$ \\
			\hline
		\end{tabular}
	\end{center}
	\caption{Confusion matrix of a binary classification task. Between square brackets, the related instance-dependent classification costs for transfer fraud are given.}
	\label{tab:confusion_matrix_general}
\end{table}
\egroup
The outcome of a classification task is usually related to costs for incorrect classifications and benefits for correctly classified instances. Let $C_i(\yh|y)$ be the cost of predicting class $\yh$ for an instance $i$ when the true class is $y$ (i.e. $y, \yh\in\{0,1\}$). If $\yh=y$ then the prediction is correct, while if $\yh\neq y$ the prediction is incorrect. In general, the costs can be different for each of the four cells in the confusion matrix and can even be instance-dependent, in other words, specific to each transaction $i$ as indicated in Table \ref{tab:confusion_matrix_general}. \cite{hand2008performance} proposed a cost matrix, where in the case of a false positive (i.e. incorrectly predicting a transaction as fraudulent) the associated cost is the administrative cost $C_i(1|0)=c_f$. This fixed cost $c_f$ has to do with investigating the transaction and contacting the card holder. When detecting a fraudulent transfer, the same cost $C_i(1|1)$ is allocated to a true positive, because in this situation, the card owner will still need to be contacted. In other words, the action undertaken by the company towards an individual transaction $i$ comes at a fixed cost $c_f\geq0$, regardless of the nature of the transaction. However, in the case of a false negative, in which a fraudulent transfer is not detected, the cost is defined to be the amount $C_i(0|1)=A_i$ of the transaction $i$. The instance-dependent costs are summarized in Table \ref{tab:confusion_matrix_general}. We argue that the proposed cost matrix in Table \ref{tab:confusion_matrix_general} is a reasonable assumption. However, the framework that is presented in this paper, including the algorithms cslogit and csboost, can deal with any cost matrix. For example, rather than using a fixed cost for false positives, one could choose to incorporate a variable cost that reflects the level of friction that the card holder experiences.

\section{Making optimal cost-based decisions}

Given the cost specification for correct and incorrect predictions, an instance should be predicted to have the class leading to the smallest \textit{expected} loss \citep{elkan2001foundations}. Here the expectation is calculated using the conditional probability of each class given the instance. These conditional probabilities are estimated by a classification algorithm. In general, a classification algorithm models the relation between $d$ explanatory variables $\boldsymbol{X} = \left(X_1, \ldots,X_d\right)$ and the binary response variable $Y\in\{0,1\}$. Such a model can be used to predict the fraud propensity of transactions on the basis of their observed variables $\xb\in\boldsymbol{X}$. In particular, a classification algorithm is a function that models the conditional expected value of $Y$:
\begin{equation*}
s:\boldsymbol{X}\rightarrow [0,1] : \xb \mapsto s(\xb)=E\left(Y|\xb\right)=P\left(Y=1|\xb\right).
\end{equation*}
Thus, a classification algorithm provides a continuous score $s_i:=s(\xb_i)\in[0,1]$ for each transaction $i$. This score $s_i$ is a function of the  observed features $\boldsymbol{x}_i$ of transaction $i$ and presents the fraud propensity of that transaction. Here we assume that legitimate transfers (class $0$) have a lower score than fraudulent ones (class $1$).

The optimal cost-based prediction for transaction $i$ is the class $\yh$ that minimizes its \textit{expected loss},
\begin{align}\label{eq:general_loss_function}
\begin{split}
EL(\xb_i, \yh) &= \sum_{y}P(Y=y|\xb_i)C_i(\yh|y) \\
&= P(Y=0|\xb_i)C_i(\yh|0) + P(Y=1|\xb_i)C_i(\yh|1)
\end{split}
\end{align}
The role of a classification algorithm is to estimate the probability $P(Y=y|\xb_i)$ for each transaction $i$ where $y$ is the true class of the transaction. The optimal prediction for a transaction is class $1$ (fraud) if and only if the expected loss of this prediction is less than the expected loss of predicting class $0$ (legitimate), i.e. if and only if
\begin{align*}
& EL(\xb_i, \yh=1) < EL(\xb_i, \yh=0) \\
\Leftrightarrow \text{ }& P(Y=0|\xb_i)C_i(1|0) + P(Y=1|\xb_i)C_i(1|1) < P(Y=0|\xb_i)C_i(0|0) + P(Y=1|\xb_i)C_i(0|1) \\ 
\Leftrightarrow \text{ }& (1-s_i)C_i(1|0)+s_iC_i(1|1) < (1-s_i)C_i(0|0)+s_iC_i(0|1) \\
\Leftrightarrow \text{ }& s_i > \frac{C_i(1|0) - C_i(0|0)}{C_i(1|0) - C_i(0|0) + C_i(0|1) - C_i(1|1)}
\end{align*}
given $s_i=P(Y=1|\xb_i)$. Thus, using the costs for transfer fraud as indicated in Table \ref{tab:confusion_matrix_general}, the threshold for making the optimal decision for a transaction $i$ is
\begin{equation}\label{eq:optimal_threshold}
t^*_i = \frac{C_i(1|0) - C_i(0|0)}{C_i(1|0) - C_i(0|0) + C_i(0|1) - C_i(1|1)}=\frac{c_f}{A_i}
\end{equation}
assuming that the transferred amount $A_i$ is nonzero. In conclusion, the optimal prediction for transaction $i$ with score $s_i$ is class $1$ (fraud) if and only if $s_i > t^*_i=c_f/A_i$, while transfer $i$ is predicted as legitimate if and only if $s_i \leq t^*_i =c_f/A_i$.
Making the prediction $\yh$ for a transaction implies acting as if $\yh$ is the true class of that transaction. Note that the $s_i$,  as estimated by the classification model, are assumed to be true calibrated probabilities rather than just scores that rank the transfers from most suspicious to least. This is especially important when making a decision based on these probabilities and their respective theshold.

The essence of cost-sensitive decision making is that, even when some class is more probable, it can be more profitable to act as if another class is true. For example, it can be rational to block a large transaction even if the transaction is most likely legitimate. Consider, for example, a transaction of \euro 1,000 with a small estimated fraud propensity of $10\%$. If the transaction is classified as legitimate, the expected loss is \euro 100 as, on average, $1$ in $10$ of these kind of transactions is in fact expected to be fraudulent. On the other hand, if the transaction is classified as fraudulent, the expected loss is the administrative cost $c_f$, for example \euro 10. Therefore, this transaction will be treated as fraudulent despite its fraud probability being only $10\%$. While it may be of interest to a financial institution to minimize false positives, the ultimate goal of the company is to maximize profits which is better addressed by the minimization of the financial costs. By using a fraud detection system, the financial institution will be able to identify fraudulent transfers and thus prevent money from being stolen from its customers, hereby reducing losses and thus generating a profit as compared to accepting or rejecting all transactions.

\section{Cost of a fraud detection model}

Let $\mathcal{D}$ denote a set of $N$ transactions which consists of the observed predictor-response pairs $\{\left(\xb_i,y_i\right)\}_{i=1}^N$, where $y_i\in\{0,1\}$ describes the binary response and $\xb_i=\left(x_{i1},\ldots, x_{id}\right)$ represents the $d$ associated predictor variables of transaction $i$. A classification model $s\left(\cdot\right)$ is trained on the set $\mathcal{D}$ such that it generates a score or fraud propensity $s_i\in[0,1]$ for each transaction $i$ based on the observed features $\xb_i$ of the transfer. The score $s_i$ is then converted to a predicted class $\yh_i\in\{0,1\}$ by comparing it with its optimal classification threshold $t^*_i$ (\ref{eq:optimal_threshold}). The cost of using $s\left(\cdot\right)$ on the transactions of $\mathcal{D}$ is calculated by \citep{bahnsen2016feature}
\begin{align}\label{eq:total_cost}
\begin{split}
Cost\left(s\left(\mathcal{D}\right)\right)=&\sum_{i=1}^{N} \bigg(y_i\Big[\yh_iC_i(1|1)+(1-\yh_i)C_i(0|1)\Big]+(1-y_i)\Big[\yh_iC_i(1|0)+(1-\yh_i)C_i(0|0)\Big]\bigg) \\
=&\sum_{i=1}^{N}y_i(1-\yh_i)A_i+\yh_ic_f
\end{split}
\end{align}
In other words, the total cost is the sum of the amounts of the undetected fraudulent transactions ($y_i=1$, $\yh_i=0$) plus the administrative cost incurred. The total cost may not always be easy to interpret because there is no reference to which the cost is compared. \citep{whitrow2009transaction}. So \cite{bahnsen2016feature} proposed the \textit{cost savings} of a classification algorithm as the cost of using the algorithm compared to using no algorithm at all. The cost of using no algorithm is
\begin{equation}\label{eq:cost_no_algorithm}
Cost_l(\mathcal{D})=\min\{Cost(s_0(\mathcal{D})),\text{ } Cost\left(s_1(\mathcal{D})\right)\}
\end{equation}
where $s_0$ refers to a classifier that predicts all the transactions in $\mathcal{D}$ as belonging to class $0$ (legitimate) and similarly $s_1$ refers to a classifier that predicts all the transfers in $\mathcal{D}$ as belonging to class 1 (fraud).

\newpage
\noindent
The cost savings is then expressed as the cost improvement of using an algorithm as compared with $Cost_l\left(\mathcal{D}\right)$,
\begin{equation}
Savings\left(s(\mathcal{D})\right)=\frac{Cost_l\left(\mathcal{D}\right) - Cost\left(s\left(\mathcal{D}\right)\right)}{Cost_l\left(\mathcal{D}\right)}
\end{equation}
In the case of credit card transaction fraud, the cost of not using an algorithm is equal to the sum of amounts of the fraudulent transactions, $Cost_l\left(\mathcal{D}\right)=\sum_{i=1}^{N}y_iA_i$. The savings are then calculated as
\begin{equation}\label{eq:savings}
	Savings\left(s\left(\mathcal{D}\right)\right)=\frac{\sum_{i=1}^{N}{y_i\yh_iA_i-\yh_ic_f}}{\sum_{i=1}^{N}y_iA_i}
\end{equation}
In other words, the costs that can be saved by using an algorithm are the sum of amounts of detected fraudulent transactions minus the administrative cost incurred in detecting them, divided by the sum of amounts of the fraudulent transactions. If $c_f=0$, then the cost savings is the proportion of amounts of fraudulent transactions that are detected.

Notice that $Cost\left(s\left(\mathcal{D}\right)\right)$ in (\ref{eq:total_cost}) depends on the optimal threshold $t^*_i$ for each transaction $i$ through the prediction $\yh_i$. The conditional expected value of $\yh_i$ is given by
\begin{equation*}
E\left[\yh_i|\xb_i\right]=P(\yh_i=1|\xb_i) \approx s\left(\xb_i\right)
\end{equation*}
Therefore, we define the \textit{average expected cost} ($AEC$) of a classification model $s(\cdot)$ on a set $\mathcal{D}$ as
\begin{align}\label{eq:average_expected_cost}
\begin{split}
AEC\left(s\left(\mathcal{D}\right)\right)
&=\frac{1}{N}E\left[Cost\left(s\left(\mathcal{D}\right)\right)\bigg|\boldsymbol{X}\right] \\
&=\frac{1}{N}\sum_{i=1}^{N} \bigg(y_i\Big[s_iC_i(1|1)+(1-s_i)C_i(0|1)\Big]+(1-y_i)\Big[s_iC_i(1|0)+(1-s_i)C_i(0|0)\Big]\bigg) \\
&=\frac{1}{N}\sum_{i=1}^{N}y_i(1-s_i)A_i+s_ic_f
\end{split}
\end{align}
Notice that the average expected cost is independent of any threshold value. Similarly, the \textit{expected savings} are computed as
\begin{align}\label{eq:expected_savings}
\begin{split}
Expected\text{ }Savings\left(s(\mathcal{D})\right)
&=E\left[Savings\left(s(\mathcal{D})\right)\bigg|\boldsymbol{X}\right] \\
&=\frac{Cost_l\left(\mathcal{D}\right) - E\left[Cost\left(s\left(\mathcal{D}\right)\right)\bigg|\boldsymbol{X}\right] }{Cost_l\left(\mathcal{D}\right)} \\
&=\frac{\sum_{i=1}^{N}{y_is_iA_i-s_ic_f}}{\sum_{i=1}^{N}y_iA_i}
\end{split}
\end{align}

\section{Cost-sensitive logistic regression and gradient tree boosting}

Popular methods for dealing with binary classification problems include logistic regression and gradient tree boosting. We opt to use logistic regression because it is widely used in the industry, it is fast to compute, easy to understand and interpret, and its flexible model structure allows for straightforward modification. Moreover, logistic regression is often used as a benchmark model to which other classification algorithms are compared. Besides logistic regression, we also adapt an algorithm for gradient tree boosting to the framework of instance-dependent costs. Decision trees are typically easy-to-use and offer high interpretability. Decision trees can also cope with complex data structures like nonlinearities and can naturally handle categorical variables. Moreover, unlike logistic regression, strongly correlated variables do not have to be removed because multicollinearity among the predictors is not an issue for tree-based methods \citep{kotsiantis2013decision}. Finally, tree boosting is a machine learning technique which is used widely by data scientists to achieve state-of-the-art results on many machine learning challenges and has been used by a series of competition winning solutions \citep{chen2016xgboost}.

\subsection{Logistic regression}

In general, a classification algorithm models the conditional mean of the binary response variable $Y$:
\begin{equation*}
s_{\thetab}:\boldsymbol{X}\rightarrow [0,1] : \xb \mapsto s_{\thetab}(\xb)=E\left(Y|\xb\right)=P\left(Y=1|\xb\right)
\end{equation*}
where $\thetab\in\Theta$ are the model parameters and $\Theta$ is the parameter space. Logistic regression is a classification model that estimates the conditional probability $P\left(Y=1|\xb\right)$ of the positive class (fraud) as the logistic sigmoid of a linear function of the feature vector $\xb$ \citep{hosmer2013applied}. The fraud propensity of a transaction is modeled as
\begin{equation}\label{eq:logistic_regression}
s_{\left(\beta_0, \betab\right)}(\xb)=P\left(Y=1|\xb\right)=\frac{1}{1+e^{-\left(\beta_0+\betab^T\xb\right)}}
\end{equation}
Here, the model parameters are $\thetab=\left(\beta_0, \betab\right)$ where $\beta_0\in\mathbb{R}$ represents the intercept and $\betab\in\mathbb{R}^d$ is the $d$-dimensional vector of regression coefficients. The problem then becomes finding the right values for the parameters that optimize a given objective (i.e. loss) function. The objective function is defined to measure the performance of a classification algorithm, given its parameters $\thetab$, on the data $\left(Y,\boldsymbol{X}\right)$:
\begin{equation*}
Q_{Y,\boldsymbol{X}}:\Theta\rightarrow\mathbb{R}:\thetab\mapsto Q_{Y,\boldsymbol{X}}(\thetab)
\end{equation*}
Usually, in the case of logistic regression, the optimal model parameters are the ones that minimize the negative binomial log-likelihood function,
\begin{equation}\label{eq:neglogbinom_objective_function}
Q^l_{Y,\boldsymbol{X}}(\beta_0, \betab)=-\frac{1}{N}\sum_{i=1}^{N}y_i\log(s_i)+(1-y_i)\log(1-s_i)
\end{equation}
where we use $s_i = s_{\left(\beta_0, \betab\right)}(\xb_i)$ to simplify the notation. Since this objection function is convex \citep{murphy2012machine}, it is generally optimized using a gradient descent algorithm like the Newton-Raphson approach. However, this objective function assigns the same weight to both false positives and false negatives. As discussed before, this is not the case in many real-world applications, including credit card transaction fraud. Instead, the weighted log-likelihood function includes a weight $w_i$ to each instance $i$ in the likelihood function depending on the instance's class (positive or negative class) or on the instance itself:
\begin{equation*}
Q^w_{Y,\boldsymbol{X}}(\beta_0, \betab)=-\frac{1}{N}\sum_{i=1}^{N}w_i\left[y_i\log(s_i)+(1-y_i)\log(1-s_i)\right]
\end{equation*}
For example, the weight assigned to an observation can be its relative cost, $w_i = c_i\big/\sum_{j=1}^{n}c_j$, where $c_i$ is the cost of misclassifying observation $i$ (e.g. $C_i(1|0)$ or $C_i(0|1)$). However, rather than optimizing a (weighted) likelihood function, the actual business objective is the minimize financial losses due to fraud and this should be reflected in the method's objective function. To that end, an instance-dependent cost-sensitive logistic model can be obtained by using the average expected cost (\ref{eq:average_expected_cost}) as objective function because it incorporates the different classification costs from Table \ref{tab:confusion_matrix_general},
\begin{equation}\label{eq:aec_objective_function}
Q^{c}_{Y,\boldsymbol{X}}\left(\beta_0, \betab\right) = AEC(\beta_0, \betab)
\end{equation}
Notice that $Q^{c}_{Y,\boldsymbol{X}}$ depends on the regression coefficients $\left(\beta_0, \betab\right)$ through the scores $s_i$.
The optimal regression parameters are the values that minimize $AEC(\beta_0, \betab)$. To find these optimal regression coefficients, the $AEC$ is minimized by using the gradient-based optimization method by \cite{kraft1988software, kraft1994algorithm}, called the sequential quadratic programming algorithm. The components of the gradient of $AEC(\beta_0, \betab)$ are given by the following partial derivatives:
\begin{align}\label{eq:gradient_aec}
\begin{split}
\frac{\partial AEC(\beta_0, \betab)}{\partial \beta_j} 
=& \frac{1}{N}\sum_{i=1}^{N} x_{ij} s_i\left(1-s_i\right) \bigg[y_i\Big(C_i(1|1) - C_i(0|1)\Big) +(1-y_i)\Big(C_i(1|0) - C_i(0|0)\Big)\bigg]  \\
=& \frac{1}{N}\sum^{N}_{i=1}x_{ij}s_i\left(1-s_i\right)\left(c_f-y_iA_i\right) \qquad (j=0, 1, \ldots, d)
\end{split}
\end{align}
where the design matrix $\boldsymbol{X}$ is defined such that its first column consists of ones, i.e. $x_{ij} = 1$ for $j = 0$. Notice that the gradient of the $AEC$ can be easily computed due to the choice of using logistic regression (\ref{eq:logistic_regression}) to model the fraud propensities $s_i$.

\begin{figure}[!t]
	\centering
	\begin{subfigure}{.5\textwidth}
		\centering
		\includegraphics[width=1\linewidth]{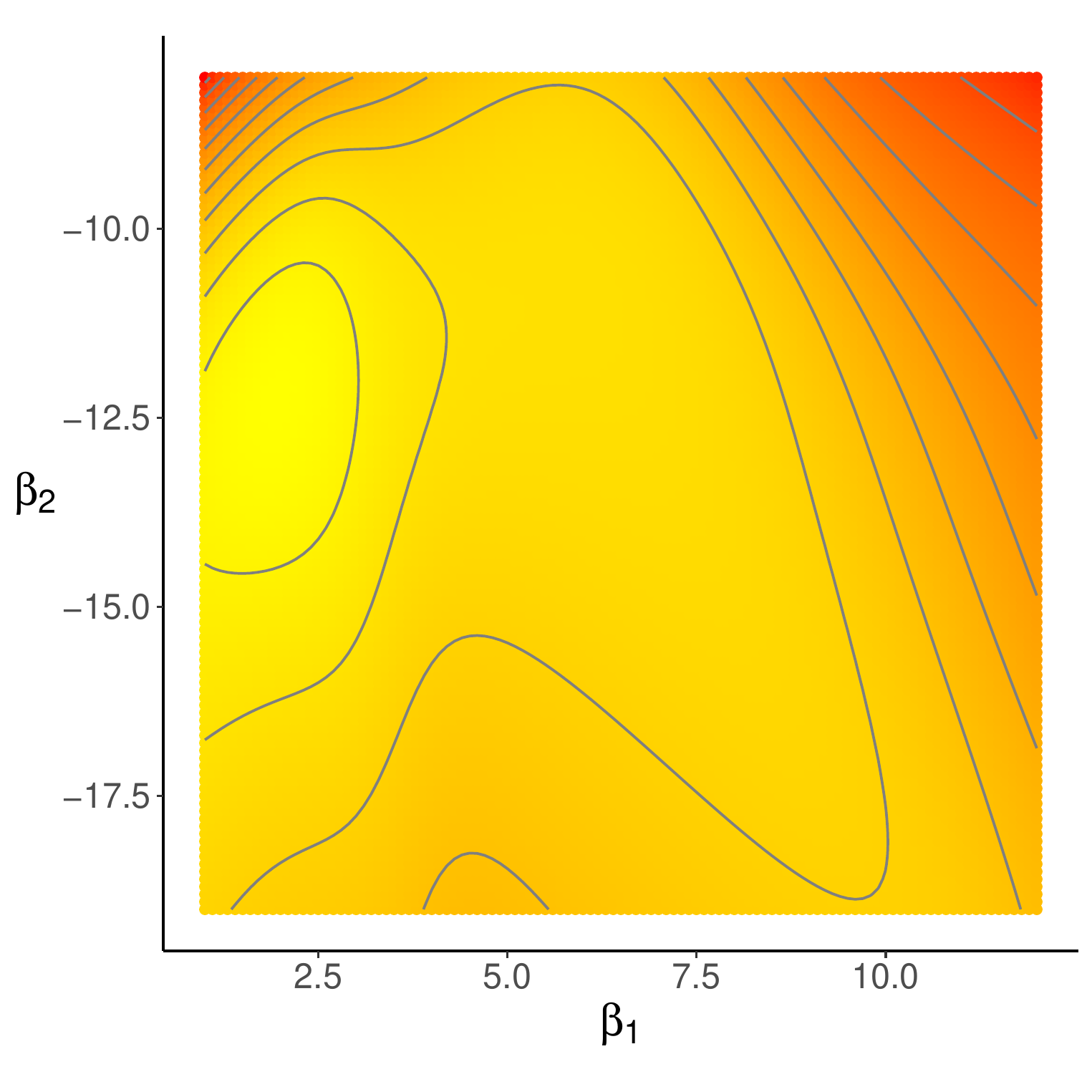}
		\caption{$\lambda = 0$}
		\label{fig:landscape_0}
	\end{subfigure}%
	\begin{subfigure}{.5\textwidth}
		\centering
		\includegraphics[width=1\linewidth]{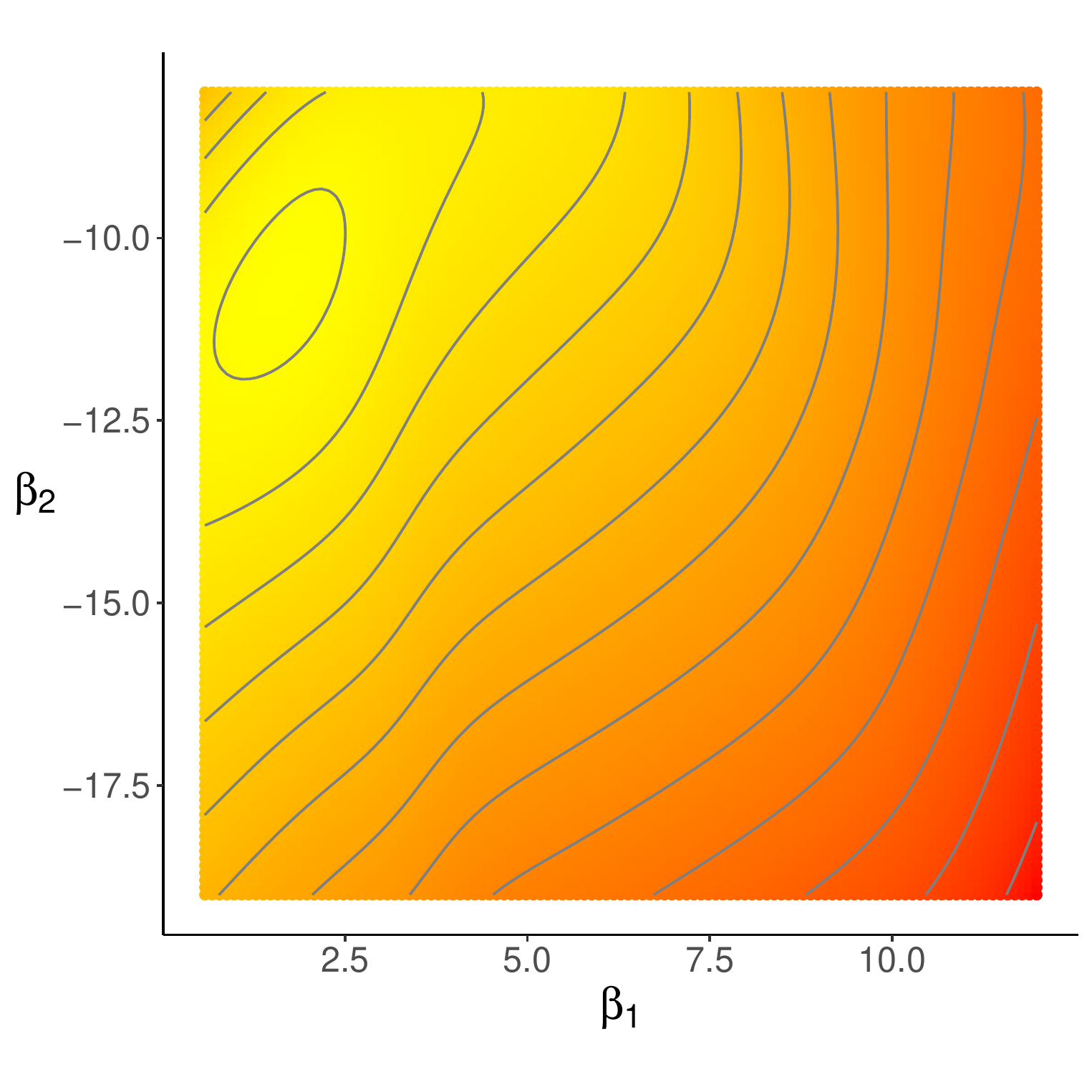}
		\caption{$\lambda>0$}
		\label{fig:landscape_lambda}
	\end{subfigure}
	\caption{Landscapes of cslogit’s objective function based on a simulated data set, where low values are in yellow and high values are in red. If the objective function only consists of the $AEC$ measure (Equation (\ref{eq:lasso_aec_objective_function}) with $\lambda = 0$) as in (a), the global minimum is unstable (due to low convexity) and multiple mimima with identical $AEC$ values may exist. On the other hand, if objective function (\ref{eq:lasso_aec_objective_function}) is augmented with the lasso penalty ($\lambda > 0$) as in (b), it induces an incline on the surface of the objective function that stabilizes the minimum and thus makes the gradient-based optimization method more efficient.}
	\label{fig:landscape}
\end{figure}

\begin{figure}[!t]
	\centering
	\includegraphics[width=0.6\textwidth]{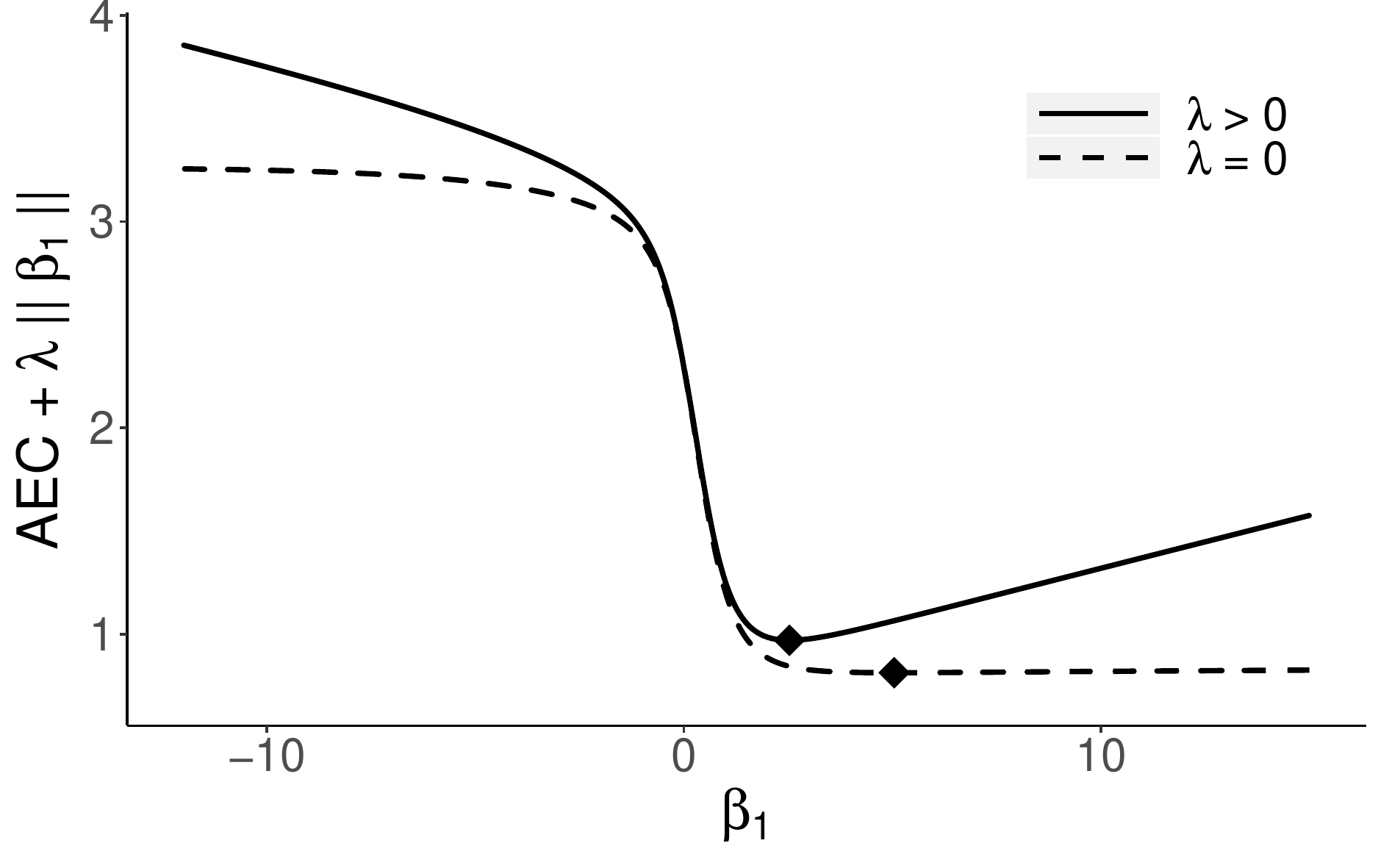}
	\caption{Objective function of cslogit, with ($\lambda > 0$) and without ($\lambda = 0$) the lasso penalty, using one regression coefficient ($\beta_1$). The optimal solution to the $AEC$ objective function ($\lambda = 0$) is very unstable, while the addition of the lasso regularization ($\lambda > 0$) stabilizes the minimum and causes the coefficient to shrink to zero.}
	\label{fig:AEC_vs_beta}
\end{figure}

In an effort to identify potential mechanisms to improve the performance, we conducted analyses of $AEC(\beta_0, \betab)$ as an objective function for logistic regression like in Figure \ref{fig:landscape_0}. We found that a pure $AEC$ objective function can exhibit multiple minima with identical $AEC$ values, and hence potentially many solutions that have the same $AEC$ value but different parameter values $(\beta_0, \betab)$ are found. Consequently, the solution that is returned by the optimization method depends on the starting values of the parameters in the training step for which we use the coefficients of a standard logistic regression model based on (\ref{eq:neglogbinom_objective_function}). Moreover, Figure \ref{fig:landscape_0} and Figure \ref{fig:AEC_vs_beta} (with $\lambda = 0$) illustrate that the optimal solution to the $AEC$ objective function is highly unstable. This means that a large shift in parameter values $(\beta_0, \betab)$ still results in $AEC$ values close to the optimum which makes it difficult for the gradient-based optimization method to converge. Therefore, we augment the objective function with a lasso penalty to avoid the undesirable behavior of finding ``unstable'' solutions. Generally, the lasso regularization penalizes model complexity and biases the gradient-based search toward simpler models as coefficients are shrinked to zero \citep{tibshirani1996regression} as shown in Figure \ref{fig:AEC_vs_beta}. Note that we do not claim that the inclusion of the lasso penalty generally results in a unique minimum \citep{stripling2018profit}.
Hence we consider the lasso-regularized version of the objective function,
\begin{equation}\label{eq:lasso_aec_objective_function}
Q^{c}_{\lambda,Y,\boldsymbol{X}}(\beta_0, \betab)=AEC(\beta_0, \betab)+\lambda ||\betab||_1
\end{equation}
where $\lambda\geq0$ is the regularization parameter and $||\betab||_1=\sum_{j=1}^{d}|\beta_j|$ is the $L_1$-norm of $\betab$. Note that the lasso regularization only penalizes the regression coefficients in $\betab$ $-$ not the intercept $\beta_0$. Clearly, the larger $\lambda$, the stronger the lasso penalty. Typically, the predictors are standardized in the lasso model so that they have zero mean (i.e. $\frac{1}{N}\sum_{i=1}^{N}x_{ij}=0$) and unit variance (i.e. $\frac{1}{N}\sum_{i=1}^{N}x_{ij}^2=1$) \citep{hastie2015statistical}. In Figure \ref{fig:landscape} and Figure \ref{fig:AEC_vs_beta}, the effect of the lasso regularization on the objective landscape is clearly visible. The inclusion of the penalty term creates an incline on the surface of the objective function that noticeably helps the gradient-based optimization method to find the minimum more efficiently.

The regularization parameter $\lambda$ cannot be directly estimated from the data and has to be determined by means of hyperparameter optimization strategies such as grid search in combination with cross-validation. The optimal value for $\lambda$ corresponds to the value with the lowest $AEC$ value. Assuming the optimal $\lambda$ value has been found, the lasso-regularized logistic regression (\ref{eq:lasso_aec_objective_function}) aims to achieve a good balance between minimizing costs and model complexity in which only predictors with sufficiently large predictive power have a nonzero regression coefficient.

\subsection{Gradient boosted decision trees}

Boosting is one of the most powerful learning concepts that has been implemented over the past twenty years. The motivation behind boosting was a process that combines the outputs of many ``weak'' classifiers to create a strong ``committee'' \citep{friedman2001elements}. A weak classifier, also called a base learner, is one whose performance is only slightly better than random guessing. An example of a weak classifier is a two terminal-node classification tree, also referred to as a tree ``stump''. In short, the aim of boosting is to apply the weak classification algorithm sequentially to repeatedly altered versions of the data, resulting in a series of weak classifiers. The predictions of all of them are then combined to produce the final prediction by a weighted majority vote. According to \cite{friedman2001elements}, trees have one element, namely inaccuracy, which prohibits them from being the perfect instrument for predictive learning. They rarely provide predictive precision similar to the best that can be accomplished with the available data. Boosting decision trees, often dramatically, enhances their precision while retaining most of their desirable data mining characteristics, like the natural handling ``mixed'' type data, being insensitive to monotone transformations of predictor variables, and having the ability to deal with irrelevant inputs. Some of the benefits of decision trees sacrificed by boosting are speed, interpretability, and potentially robustness against overlapping class distributions and particularly mislabeling of training data. A gradient-boosted model is a tree-boosting generalization that tries to mitigate these issues in order to create an precise and efficient data mining process. Gradient tree boosting is implemented in several \textsf{R} software packages, including \verb|gbm| \citep{ridgeway1999state, ridgeway2007generalized} and \verb|mboost| \citep{hothorn2006model}. However, these algorithms use an accuracy related performance measure, like the negative binomial log-likelihood (\ref{eq:neglogbinom_objective_function}), as their objective function. The \textsf{R} package \verb|xgboost|, on the other hand, is made to be extendible and allows us to easily define our own cost-sensitive objection function.

\verb|xgboost| is short for eXtreme Gradient Boosting \citep{chen2016xgboost}. It is an efficient and scalable implementation of the gradient boosting framework by \cite{friedman2000additive} and \cite{friedman2001greedy}. Both \verb|xgboost| and \verb|gbm| follow the same principle of gradient boosting, but there are some key differences in the modeling details. Specifically, \verb|xgboost| uses a more regularized model formalization to control over-fitting, which gives it better performance. The name \verb|xgboost| refers to the engineering goal to push the limit of computational resources for boosted tree algorithms. As a result, it is generally over 10 times faster than \verb|gbm| \citep{chen2016xgboost}.

Consider a data set with $N$ instances and $d$ variables $\mathcal{D}=\{(\xb_i, y_i)\}$ where $|\mathcal{D}|=N$, $\xb_i\in\mathbb{R}^d$ and $y_i\in\{0, 1\}$. A tree ensemble method uses $K$ additive functions to predict the output:
\begin{equation*}
\yh_i=\phi(\xb_i)=\sum_{k=1}^{K}f_k(\xb_i),\quad f_k\in\mathcal{F},
\end{equation*}
where $\mathcal{F}=\{f(\xb)=w_q(\xb)\text{ }|\text{ }q:\mathbb{R}^d\rightarrow T, w\in \mathbb{R}^T\}$ is the space of decision trees. Here $q$ represents the structure of each tree that maps an instance to the corresponding leaf index. $T$ is the number of leaves in the tree. Each $f_k$ corresponds to an independent tree structure $q$ and leaf weights $w$. For a given instance, the decision rules in the trees (given by $q$) are used to classify it into the leaves and calculate the final prediction by summing up the weights in the corresponding leaves (given by $w$). To learn the set of functions used in the model, the following \textit{regularized} objective is optimized:
\begin{align}\label{eq:regularized_gradient_boosting_objective}
\begin{split}
\mathcal{L}(\phi)=\sum_i l\left(\yh_i, y_i\right) + \sum_k \Omega(f_k) \\
\text{where } \Omega(f)=\gamma T+\frac{1}{2}\lambda ||w||^2
\end{split}
\end{align}
Here $l$ is a differentiable convex objective (i.e. loss) function that measures the difference between the prediction $\yh_i$ and the target $y_i$. The second term $\Omega$ penalizes the complexity of the model (i.e., the decision tree functions). The additional regularization term helps to smooth the final learnt weights to avoid over-fitting. Intuitively, the regularized objective tends to select a model employing simple and predictive functions. When the regularization parameter is set to zero, the objective falls back to the traditional gradient tree boosting. The tree ensemble method (\ref{eq:regularized_gradient_boosting_objective}) includes functions as parameters and cannot be optimized using traditional optimization methods in Euclidean space. Instead, the model is trained in an additive manner. Formally, let $\yh_i^{(t)}$ be the prediction of the $i$-th instance at the $t$-th iteration, then $f_t$ is added to minimize the following objective:
\begin{equation*}
\mathcal{L}^{(t)}=\sum_{i=1}^{N}l\left(y_i, \yh_i^{(t-1)}+f_t(\xb_i) \right) + \Omega(f_t)
\end{equation*}
This means that the $f_t$ that most improves the model according to (\ref{eq:regularized_gradient_boosting_objective}) is greedily added. Second-order approximation can be used to quickly optimize the objective in the general setting \citep{friedman2000additive}:
\begin{equation*}
\mathcal{L}^{(t)}\simeq\sum_{i=1}^{N}\bigg[l\left(y_i, \yh_i^{(t-1)}\right)+g_if_t(\xb_i)+\frac{1}{2}h_if_t^2(\xb_i)\bigg] + \Omega(f_t)
\end{equation*}
where
\begin{equation*}
g_i=\partial_{\yh^{(t-1)}}l\left(y_i, \yh_i^{(t-1)}\right) \text{ and } h_i=\partial^2_{\yh^{(t-1)}}l\left(y_i, \yh_i^{(t-1)}\right)
\end{equation*}
are first and second order gradient statistics on the objective function. The constant terms can be removed to obtain the following simplified objective at step $t$:
\begin{equation}\label{eq:simplified_xgboost_objective}
\mathcal{\tilde{L}}^{(t)}\simeq\sum_{i=1}^{N}\bigg[g_if_t(\xb_i)+\frac{1}{2}h_if_t^2(\xb_i)\bigg] + \Omega(f_t)
\end{equation}
Notice that this last equation only contains the gradient and second order gradient of objective function $l$. In order to achieve an algorithm for instance-dependent cost-sensitive gradient tree boosting, we substitute the gradient and second order gradient of the \textit{average expected cost} (\ref{eq:average_expected_cost}). Given the provided training data $(Y, \boldsymbol{X})$ and a set of predictions $\eta_i$ ($i=1, \ldots, N$) before logistic transformation,  the probability scores are defined as
\begin{equation*}
s_i=\frac{1}{1+e^{-\eta_i}}\qquad (i=1, \ldots, N)
\end{equation*}
The gradient and second order gradient of the $AEC$ (\ref{eq:average_expected_cost}) are then $(i=1, \ldots, N)$:
\begin{align}
\begin{split}
g_i=\frac{\partial AEC}{\partial \eta_i} &= s_i\left(1-s_i\right) \bigg[y_i\Big(C_i(1|1) - C_i(0|1)\Big) +(1-y_i)\Big(C_i(1|0) - C_i(0|0)\Big)\bigg]  \\
&=s_i\left(1-s_i\right)\left(c_f-y_iA_i\right) \\
h_i = \frac{\partial^2 AEC}{\partial \eta_i^2} &= s_i\left(1-s_i\right)\left(1-2s_i\right) \bigg[y_i\Big(C_i(1|1) - C_i(0|1)\Big) +(1-y_i)\Big(C_i(1|0) - C_i(0|0)\Big)\bigg]  \\
&= s_i\left(1-s_i\right)\left(1-2s_i\right)\left(c_f-y_iA_i\right) \\
&= \frac{\partial AEC}{\partial \eta_i} (1-2s_i)
\end{split}
\end{align}

\section{The cslogit and csboost algorithms}

The pseudocode for the cslogit algorithm is provided in Algorithm 1. The goal of the algorithm is to find estimates for the regression coefficients $(\beta_0, \betab)$ that minimize the cost-sensitive objective function $Q^{c}_{\lambda,Y,\boldsymbol{X}}(\beta_0, \betab)=AEC(\beta_0, \betab)+\lambda ||\betab||_1$.  The algorithm starts by fitting a regular logistic regression model (\ref{eq:neglogbinom_objective_function}) to the provided data set to obtain initial values $(\beta_0^0, \betab^0)$ for the optimization. The cslogit algorithm uses the gradient based-optimization method from \cite{kraft1988software, kraft1994algorithm} to sequentially minimize the regularized average expected cost. At each iteration $m$, the current estimate $(\beta_0^m, \betab^m)$ is improved by using the gradient of $Q^{c}_{\lambda,Y,\boldsymbol{X}}$ (\ref{eq:gradient_aec}) evaluated at $(\beta_0^m, \betab^m)$. The algorithm terminates when the objective value converges or all of the regression coefficients converge. The details of the termination criteria are given in Algorithm 1. If convergence does not occur, the algorithm terminates after a user-specified number of iterations, which is set at 10,000 by default. The result of the cslogit algorithm is the set of regression parameters $(\hat{\beta}_0, \boldsymbol{\hat{\beta}})$ with the lowest value for $Q^{c}_{\lambda,Y,\boldsymbol{X}}$.

The csboost algorithm is essentially a wrapper for the \verb|xgb.train| function from the \verb|xgboost| package in \textsf{R}. The details are discussed below. The \verb|xgboost| implementation allows us to specify the average expected cost (\ref{eq:average_expected_cost}) as both the objective function and evaluation metric. In the case of csboost, the training dataset is specified using a formula and a data frame as is most common in \textsf{R} implementations, rather than using an \verb|xgb.DMatrix| as is the case with \verb|xgb.train|. This makes the csboost function more user-friendly.

\bgroup
\def\arraystretch{1.1}
\setlength{\tabcolsep}{5pt}
\begin{table}[t!]
	\centering\small
	\begin{tabular}{l}
		\hline
		\textbf{Algorithm 1: cslogit}\\ \hline
		\textbf{Inputs}\\
		\quad \tabitem $\{(\xb_i,y_i)\}^N_{i=1}$, data set with $d$ predictors;  \\
		\quad \tabitem $\boldsymbol{C}\in\mathbb{R}^{N\times2}$, cost matrix. For each instance, the first (resp. second) column contains\\
		\qquad \text{ } the cost of correctly (resp. wrongly) predicting the binary class of the instance;  \\
		\quad \tabitem $\lambda$, regularization parameter;\\
		\quad \tabitem $M$, maximum number of iterations (default is 10,000);\\
		\quad \tabitem $\Delta f$, relative tolerence of the objective function (\ref{eq:lasso_aec_objective_function}) (default is $1\mathrm{e}{-8}$); \\
		\quad \tabitem $\Delta \beta$, relative tolerence of the regression coefficients (default is $1\mathrm{e}{-5}$); \\
		\textbf{Starting values for the regression parameters} \\
		\quad Fit a regular logistic regression model (\ref{eq:neglogbinom_objective_function}) to the data to obtain initial starting values $(\beta_0^0, \betab^0)$ for\\
		\quad the optimization. \\
		\textbf{The main loop of the algorithm} uses the gradient based-optimization method from\\
		\quad \cite{kraft1988software, kraft1994algorithm} to sequentially minimize $Q^{c}_{\lambda,Y,\boldsymbol{X}}(\beta_0, \betab)=AEC(\beta_0, \betab)+\lambda ||\betab||_1$. \\
		\quad At each iteration $m\leq M$, the current estimate $(\beta_0^m, \betab^m)$ is improved by minimizing $Q^{c}_{\lambda,Y,\boldsymbol{X}}$ using\\
		\quad the gradient (\ref{eq:gradient_aec}) evaluated at $(\beta_0^m, \betab^m)$. \\
		\textbf{Termination criteria} \\
		\quad The algorithm terminates when $|Q^{c}_{\lambda,Y,\boldsymbol{X}}(\beta_0^{m}, \betab^m) - Q^{c}_{\lambda,Y,\boldsymbol{X}}(\beta_0^{m-1}, \betab^{m-1})|< \Delta f \cdot Q^{c}_{\lambda,Y,\boldsymbol{X}}(\beta_0^{m}, \betab^m)$ \\
		\quad or when $|\beta_j^m-\beta_j^{m-1}|<\Delta \beta \cdot \beta_j^m$ for all $j=0, 1, \ldots, d$.\\
		\quad If the algorithm does not converge, it will terminate when the maximum number of iterations is\\
		\quad reached in which case a warning message is written to the command line.\\
		\textbf{Output}\\
		\quad Finally, the set of regression parameters $(\hat{\beta}_0, \boldsymbol{\hat{\beta}})$ with the lowest objective value $Q^{c}_{\lambda,Y,\boldsymbol{X}}$ is returned.\\
		\hline
	\end{tabular}
\end{table}
\egroup

\newpage
The functions cslogit and csboost are the main function of their respective packages with the same name, \verb|cslogit| and \verb|csboost|. Both packages are written in \textsf{R} and are available at \url{github.com/SebastiaanHoppner/CostSensitiveLearning}.

\subsection{User interface of cslogit}

\begin{table}[!b]
	\begin{center}
		\begin{tabular}{| l | l |}\hline
			Parameter & Description  \\ \hline\hline
			\texttt{maxeval} & maximum number of iterations (default is 10,000).\\ \hline
			\texttt{ftol\_rel}& obtain the minimum of the objective function to within a relative tolerance \\
			& (default is $10^{-8}$). \\ \hline
			\texttt{xtol\_rel}& obtain optimal regression coefficients to within a relative tolerance \\
			& (default is $10^{-5}$).  \\\hline
			\texttt{start}& starting values for the coefficients in the linear predictor.  By default a logistic \\
			& regression model is fitted in order to use the coefficients as starting values. \\ \hline
			\texttt{lb}& vector with lower bounds of the coefficients. By default the intercept is \\
			&  unbounded and the other coefficients have a lower bound of \\
			& \verb|-max(50, abs(options$start[-1]))|. \\ \hline
			\texttt{ub}&  vector with upper bounds of the coefficients. By default the intercept is \\
			& unbounded and the other coefficients have an upper bound of\\
			& \verb|max(50, abs(options$start[-1]))|.\\ \hline
			\texttt{check\_data}& should the data be checked for missing or infinite values (default is \texttt{TRUE})\\
			& or not (\texttt{FALSE}).  \\ \hline
			\texttt{print\_level}& controls how much output is shown during the optimization process.\\
			& Possible values: \\
			& 0 (default) \quad no output \\
			& 1 \qquad\qquad\text{ }\text{ }\text{ } show iteration number and value of objective function (\ref{eq:lasso_aec_objective_function}) \\
			& 2 \qquad\qquad\text{ }\text{ }\text{ } $1+$ show value of coefficients \\ \hline
		\end{tabular}
	\end{center}
	\caption{Control parameters of cslogit.}
	\label{tab:control_parameters_cslogit}
\end{table}

The principal function of the \verb|cslogit| package is the function \verb|cslogit()| taking arguments
\[\verb|cslogit(formula, data, cost_matrix, lambda, options = list())|\]
where \verb|formula| is a symbolic description of the model to be fitted, e.g. via \verb|formula = y ~ x1 + x2|, which must include an intercept, and \verb|data| is a data frame containing the variables in the model. Parameter \verb|lambda| is the value that controls the lasso regularization of the regression coefficients in (\ref{eq:lasso_aec_objective_function}). Argument \verb|cost_matrix| is a matrix in $\mathbb{R}^{n\times2}$. For each instance, its first (resp. second) column contains the cost of correctly (resp. incorrectly) predicting the binary class of the instance. Given the costs in Table \ref{tab:confusion_matrix_general}, \verb|cost_matrix| can be implemented as
\begin{align*}
	&\texttt{cost\_matrix <- matrix(nrow = nrow(data), ncol = 2)}\\
	&\texttt{cost\_matrix[, 1] <- ifelse(data\$y == 1, fixed\_cost, 0)}\\
	&\texttt{cost\_matrix[, 2] <- ifelse(data\$y == 1, transferred\_amount, fixed\_cost)}
\end{align*}
Finally, \verb|options| comprises a list of control parameters as described in Table \ref{tab:control_parameters_cslogit}. The logistic regression model fitted by cslogit is a list of class `\verb|cslogit|'. The methods inherited in this way include \verb|summary()| and \verb|plot()| functions to display the evolution of the objective value and regression parameters, as well as a \verb|predict()| function to compute estimated probabilities for unseen data.

Futhermore, the \verb|cslogit| package contains the function \verb|cv.cslogit()| which performs a cross-validation procedure for the \verb|lambda| parameter of cslogit. It takes the same arguments as the cslogit function, but needs a whole sequence of possible values for \verb|lambda| rather than a single value. Additionally, one must also specify the number of folds through the \verb|nfolds| parameter which has a default value of $10$.

\subsection{User interface of csboost}

\begin{table}[!b]
	\begin{center}
		\begin{tabular}{| l | l |}\hline
			Parameter & Description  \\ \hline\hline
			\texttt{nrounds}& maximum number of boosting iterations. \\ \hline
			\texttt{params}& the list of parameters for the \verb|xgboost| algorithm.\\
			& The complete list of parameters can be consulted in the documen-\\
			& tation \verb|help(xgb.train)|. \\\hline
			\texttt{verbose} & If 0, \verb|xgboost| will stay silent. If 1, it will print information about \\
			& performance. If 2, some additional information will be printed out. \\ \hline
			\texttt{print\_every\_n} & Print each n-th iteration evaluation messages when \texttt{verbose > 0}. \\
			& Default is 1 which means all messages are printed. \\  \hline
			\texttt{early\_stopping\_rounds} & If \verb|NULL|, the early stopping function is not triggered. If set to an \\
			& integer $k$, and argument \verb|test| is specified, training with the valida-\\
			& tion set will stop if the performance does not improve for $k$ rounds. \\ \hline
		\end{tabular}
	\end{center}
	\caption{Input arguments of csboost.}
	\label{tab:arguments_csboost}
\end{table}

The main function of the \verb|csboost| package is the function \verb|csboost()| which is essentially a wrapper for the function \verb|xgb.train()| from the \verb|xgboost| package:
\begin{align*}
\texttt{csboost(}&\texttt{formula, train, test = NULL, cost\_matrix\_train, cost\_matrix\_test = NULL,}\\
&\texttt{nrounds, params = list(),verbose = 1, print\_every\_n = 1L, }\\
&\texttt{early\_stopping\_rounds = NULL)}
\end{align*}
where \verb|formula| and \verb|train| specify the training data in the usual way. One can also specify a validation set through the \verb|test| argument. Similar to cslogit, \verb|cost_matrix_train| is the instance-dependent cost matrix for the training set. If \verb|test| is provided, one must also specify \verb|cost_matrix_test|. The remaining arguments are described in Table \ref{tab:arguments_csboost}. The output of csboost is a list of class `\verb|csboost|' which contains the fitted gradient boosted tree model. The accompanying functions include \verb|summary()|, \verb|plot()| and \verb|predict()|, similar to the functions linked to the `\verb|cslogit|' class.

\section{Experiments}

We assess the performance of our new methods by benchmarking them against their cost-insentitive counterparts, (regular) logistic regression and the xgboost algorithm. To do so, we apply the classification techniques to a publicly available real-life fraud data set and a data set provided by a large bank. All methods are evaluated using Savings, Expected Savings, Precision, Recall and $F_1$ measure where we use the instance-dependent thresholds (\ref{eq:optimal_threshold}).

\subsection{Data sets}

The first data set is the Credit Card Transaction Data available at \url{kaggle.com/mlg-ulb/creditcardfraud}. The data consists of transactions made by credit cards in September 2013 by European cardholders. This data set presents transactions that occurred in two days, where we have $492$ frauds out of 284,807 transactions. The data set is highly unbalanced, the positive class (frauds) account for $0.172\%$ of all transactions. It contains only numerical input variables which are the result of a PCA transformation. Due to confidentiality issues, the original features and more background information about the data cannot be provided. Features V1, V2, ..., V28 are the principal components obtained with PCA. The only features which have not been transformed with PCA are `Time' and `Amount'. Feature `Time' contains the seconds elapsed between each transaction and the first transaction in the data set. The feature `Amount' is the transaction amount where we remove a few transactions with a zero amount. Feature `Class' is the response variable which takes value $1$ in case of fraud and $0$ otherwise. We select the features V1, V2, ..., V28 and the logarithmically transformed Amount as predictor variables for the classification methods. Each of these 29 predictors are scaled to zero mean and unit variance.  

The second data set has been provided to our research group by a large bank and will be refered to as the Bank data set. The data contains 31,763 transactions made between September 2018 and July 2019. In total there are 506 (1.6\%) transactions labeled as fraudulent. The data set contains 21 numerical input features, 3 categorical features and the fraud indictor as response variable. 

In each data set, we scale the (non-categorical) predictors to zero mean and unit variance. To keep the analysis of the data sets manageable, we only consider main effects and we do not include interactions of any degree.

\subsection{Experimental design}

For each data set, we perform $5$ replications of two-fold cross validation. For the experiments, we want to ensure that both folds contain an equal balance of high, middle and low value fraud cases. Therefore, we compute the $33\%$ and $66\%$ quantiles of the fraudulent amounts, and we divide every transfer in one of three categories (i.e. high, middle or low) depending on their amount with respect to these quantiles. For the Credit Card Transaction Data, transfers with an amount below \euro$1.10$ are categorized as ``'low'', transfers between \euro$1.10$ and \euro$99.99$ are considered ``middle'', and transfers above \euro$99.99$ are categorized as ``high''. Similarly for the Bank data set, the $33\%$ and $66\%$ quantiles of the fraudulent amounts are \euro$1999.33$ and \euro$5000$, respectively. Using these three categories, each fold in the cross validation procedure is stratified according to the binary response variable  as well as the amount category in order to obtain similar distributions in the folds as observed in the original data set.

\subsection{Results}

\begin{figure}[!t]
	\centering
	\includegraphics[width=1\textwidth]{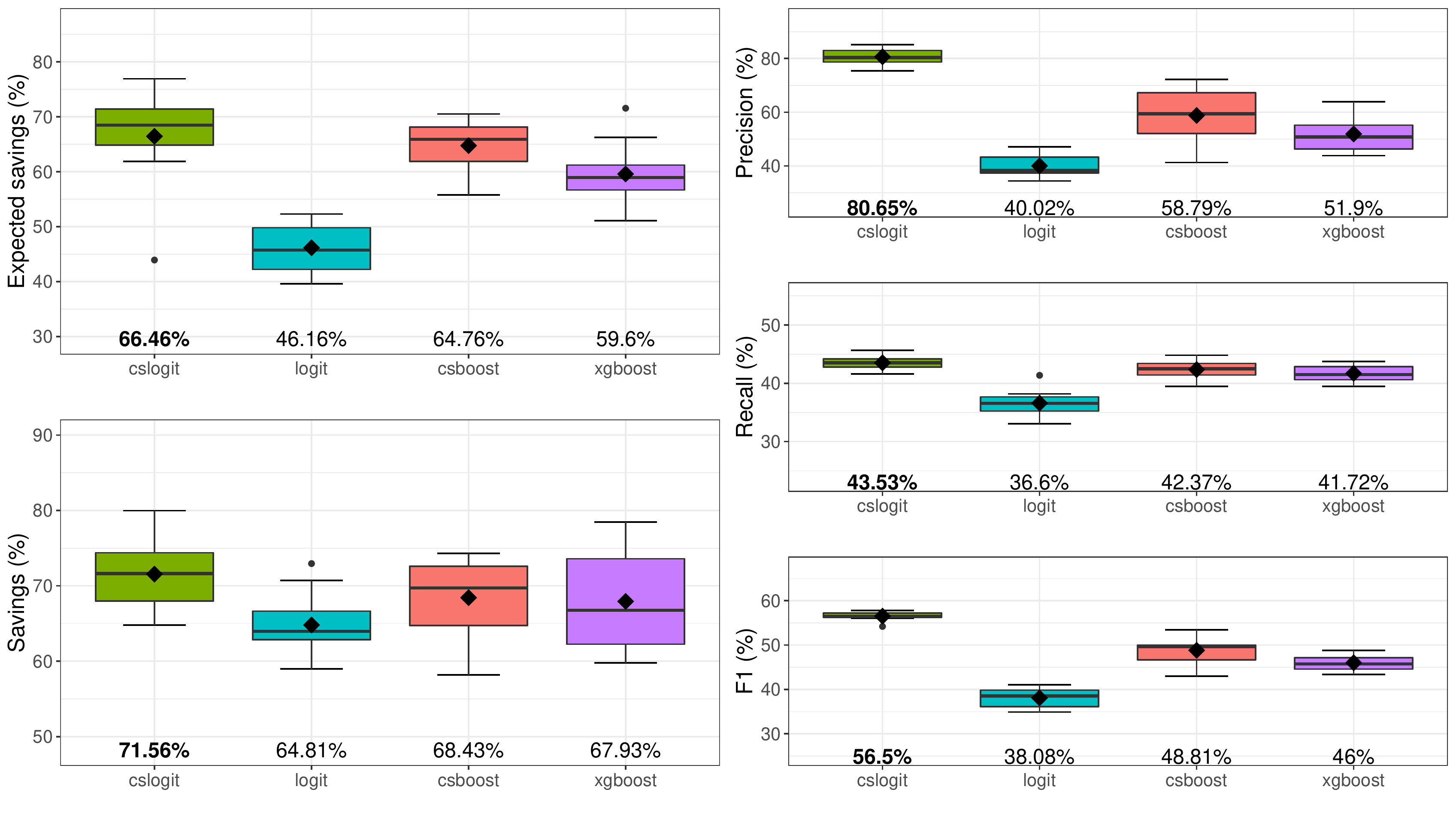}
	\caption{Five times two-fold cross validation results for the credit card transaction data.}
	\label{fig:results_creditcard}
\end{figure}

Figure \ref{fig:results_creditcard} contains the results of the $5\times2$-fold cross validation procedure for the Credit Card Transaction Data. In terms of expected savings, cslogit outperforms logistic regression and csboost outperforms gradient boosted trees (xgboost) as can be seen in the top left figure. Of course, this is expected since both cost-sensitive methods have the average expected cost (AEC) in their objective function which they minimize. Note that by minimizing the AEC (\ref{eq:average_expected_cost}), the expected savings measure is maximized (\ref{eq:expected_savings}). After applying the instance-dependent cost-related threshold (\ref{eq:optimal_threshold}), the savings measure can be assessed. Both cost-sensitve methods outperform their classical counterparts although the difference between them is smaller. Although cslogit and csboost are designed to optimize cost-related measures, like AEC and expected savings, it is interesting to notice that cslogit and csboost do better than logistic regression and xgboost in terms of precision, recall and $F_1$ on this data set. Overall, the difference in performance between cslogit and logit is larger than the difference between csboost and xgboost. This is also reflected in their cost: on average across the 10 folds, the difference in total costs between cslogit and logit is \euro$2005.83$ while the difference in total costs between csboost and xgboost is \euro$169.44$. The average execution times are reported in Table \ref{tab:times_creditcard}. These execution times were measured on an Intel core i5 with 2.7 GHz and 8 GB RAM.
\begin{table}[!b]
	\begin{center}
	\begin{tabular}{c c c c c}
		\hline
		& cslogit & logit & csboost & xgboost  \\ \hline
		time & 4.82 & 2.94 & 5.65 & 9.48 \\ \hline
	\end{tabular}
	\end{center}
\caption{Average time (in seconds) to fit each of the classification methods on the training set of the credit card transaction data.}
\label{tab:times_creditcard}
\end{table}

\begin{figure}[!t]
	\centering
	\includegraphics[width=1\textwidth]{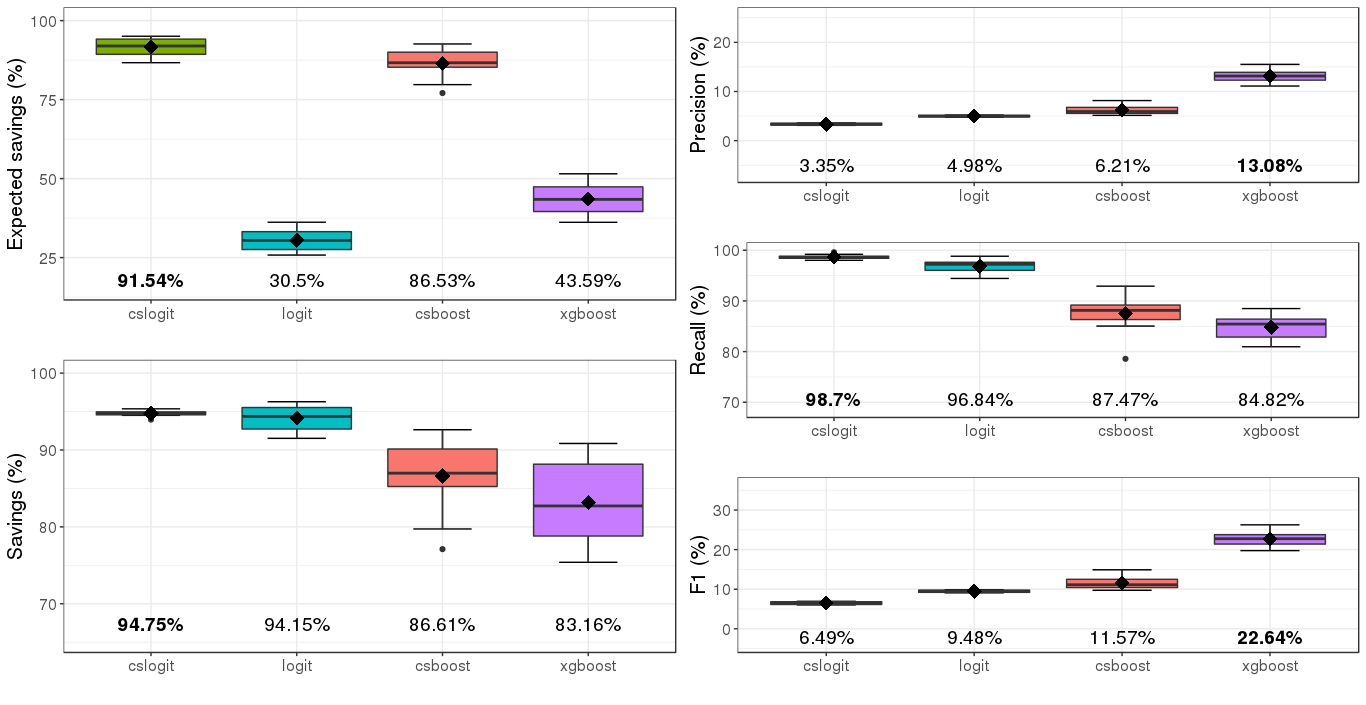}
	\caption{Five times two-fold cross validation results for the bank transaction data.}
	\label{fig:results_bank}
\end{figure}

Figure \ref{fig:results_bank} contains the results of the $5\times2$-fold cross validation procedure for the Bank data set. As expected, cslogit and csboost largely outperform logit and xgboost in terms of expected savings. When applying the instance-dependent cost-related threshold (\ref{eq:optimal_threshold}), the predicted fraud probabilities are converted into binary decisions (i.e. fraud or not), on which the savings measure is computed. Compared to the previous data set, cslogit slightly outperforms logistic regression while the difference between csboost and xgboost is larger on average. The average difference between cslogit and logit in terms of average cost for a single transfer between is \euro$0.62$. Although this difference might seem small, it can accumulate to large amounts if we take into account that a bank may process over 100,000 transactions each day. The average differnce between csboost and xgboost is terms of average cost for a single transaction is \euro$3.25$.

cslogit and csboost are designed to minimize the financial losses due to fraud by taking into account the various costs between transactions due to classification. The results in Figure \ref{fig:results_bank} illustrate that model selection purely based on accuracy related performance measures, such as precision, recall and $F_1$, likely results in models that have a higher cost. This is clearly demonstrated by the fact that cslogit has the highest cost savings while simultaneously has the worst $F_1$ values.

\section{Conclusions and future research}

In this paper, we presented two new classifiers called cslogit and csboost which are based on lasso-regularized logistic regression and gradient tree boosting, respectively. Each method directly minimizes the proposed instance-dependent cost measure in the model construction step. As a result, cslogit and csboost aim to create the detection model which minimizes the financial loss due to fraud. Furthermore, based on the instance-dependent cost matrix for transfer fraud, we derived a transfer-specific threshold that allows for making the optimal cost-based decision for each transaction.

In our benchmark study, cslogit and csboost outperform their classical counterpart models, which are ignorant of any classification costs, in terms of the costs saved due to detecting fraud. We conclude that model selection based on accuracy related measures, such as precision and recall, leads to more costly results. In this paper, we have shown that our proposed methods align best with the core business objection of cost minimization by prioritizing the detection of high-amount fraudulent transfers.

Concerning future research, we intend to include artificial neural networks to the collection of instance-dependent cost-sensitive classifiers. This method, called csnet, computes the gradient of the proposed instance-dependent cost measure in its backpropagation algorithm. An extensive empirical evaluation and comparison between the cost-sensitive methods cslogit, csboost and csnet can then be conducted. Although cslogit and csboost are used for detecting card transaction fraud, the framework that is presented in this paper, including both methods, can deal with any cost matrix. Therefore, cslogit and csboost have potential in multiple fraud detection domains such as insurance fraud \citep{dionne2009optimal}, e-commerce fraud \citep{nanduri2020microsoft} and social security fraud \citep{van2017gotcha}, as well as other analytical tasks where costs are important such as credit-risk evaluation \citep{baesens2003using, verbraken2014development} and customer churn prediction \citep{verbeke2012}. The main adaptation to these tasks would consist of identifying the costs of classifying instances and defining the appropriate cost matrix.

\bibliographystyle{elsarticle-harv} 
\bibliography{References}

\end{document}